\documentclass[10pt,aps,prl,reprint,nofootinbib]{revtex4-1}

\usepackage{amsmath}
\usepackage{graphicx}
\usepackage[colorlinks,pdfusetitle]{hyperref}
\hypersetup{colorlinks=true,allcolors=[rgb]{1,0.56,0}}

\newcommand{\orcid}[1]{\href{https://orcid.org/#1}{#1}}
\newcommand{\e}[1]{\times10^{#1}}
\DeclareMathOperator{\sign}{sign}
\newcommand{\wh}[1]{\widehat{#1}}

\begin{document}

\title{CP-Violation with Neutrino Disappearance Alone}

\author{Peter B.~Denton}
\email{pdenton@bnl.gov}
\thanks{\orcid{0000-0002-5209-872X}}

\affiliation{High Energy Theory Group, Physics Department, Brookhaven National Laboratory, Upton, NY 11973, USA}

\begin{abstract}
The best way to probe CP violation in the lepton sector is with long-baseline accelerator neutrino experiments in the appearance mode: the appearance of $\nu_e$ in predominantly $\nu_\mu$ beams.
Here we show that it is possible to discover CP violation with disappearance experiments only, by combining JUNO for electron neutrinos and DUNE or Hyper-Kamiokande for muon neutrinos.
While the maximum sensitivity to discover CP is quite modest ($1.6\sigma$ with 6 years of JUNO and 13 years of DUNE), some values of $\delta$ may be disfavored by $>3\sigma$ depending on the true value of $\delta$.
\end{abstract}

\date{May 14, 2024} 

\maketitle

\section{Introduction}
There are three free parameters in our current model of particle physics that control the size of CP violation in their respective sector: the $\bar\theta$ term in the QCD sector which seems to be either zero or very small \cite{Pendlebury:2015lrz}, the CKM matrix \cite{Cabibbo:1963yz,Kobayashi:1973fv} describing quark mixing which is known to have some CP violation \cite{ParticleDataGroup:2018ovx,Charles:2015gya}, and the PMNS matrix \cite{Pontecorvo:1957cp,Maki:1962mu} describing lepton mixing.
It is unknown if there is CP violation in the lepton mixing matrix \cite{deSalas:2020pgw,Esteban:2020cvm,Capozzi:2021fjo,Denton:2020uda} and thus determining if CP is violated in the lepton sector is of the utmost priority in particle physics.

The best way to probe CP violation in the leptonic sector is by an appearance measurement of an oscillation maximum \cite{T2K:2001wmr,Marciano:2001tz,Dick:1999ed,Cervera:2000kp,Autiero:2003fu,Nunokawa:2007qh,Coloma:2012ji,Denton:2022een}.
That is, the detection of one flavor of neutrino in a source of neutrino that is predominantly a differently flavor at a baseline and energy that corresponds to one of the $\Delta m^2_{ij}$'s.
To date only NOvA \cite{NOvA:2016kwd} and T2K \cite{T2K:2013ppw} have strong evidence for the detection of appearance by detecting electron neutrinos in predominantly muon neutrino sources, but do not yet significantly probe CP violation past the $2\sigma$ level \cite{NOvA:2021nfi,T2K:2023smv}.
Atmospheric neutrinos, which are mostly muon neutrinos also have some evidence for appearance \cite{Super-Kamiokande:2017edb,Super-Kamiokande:2019gzr,IceCube:2019dqi}.
Since there is significant electron neutrino contribution at the source, the appearance information is thus somewhat scrambled.

We will show how it is also possible to probe CP violation, via neutrino disappearance measurements only, in three different ways: counting parameters using a specific parameterization, direct analytic calculation in a parameterization independent framework, and numerical computation.
With good measurements of the disappearance of two different flavors it is possible to determine a total of four independent parameters in the PMNS matrix.
Since the mixing matrix is only described by four parameters it is possible to learn about CP violation while only measuring CP conserving channels.
The key physics effect that makes this possible is unitarity\footnote{We note that unitarity is used for most experiment's extraction of one or more of the mixing parameters.
For example, medium-baseline reactor neutrino experiments at $L\simeq1$ km actually determine $4|U_{e3}|^2(|U_{e1}|^2+|U_{e2}|^2)$ but by unitarity this is equal to $4|U_{e3}|^2(1-|U_{e3}|^2)$ which can be expressed as $\sin^22\theta_{13}$.} and thus if there is new physics in the neutrino sector this story may get more complicated.
It is also possible to relate the amount of CP violation directly to the measured parameters from disappearance.
Given the sizable expected improvements in disappearance measurements in the $\nu_e$ channel with the Jiangmen Underground Neutrino Observatory (JUNO) and the $\nu_\mu$ channel with the Deep Underground Neutrino Experiment (DUNE) and Hyper-Kamiokande (HK), such a study is quite timely.
Moreover, disappearance has a different dependence on the oscillation parameters as well as different (and often cleaner) systematics than appearance measurements since the neutrinos at the near and far detectors are the same flavor which means that this can be a valuable cross check of CP violation probes in the appearance channel.
See also \cite{Farzan:2002ct} for a very early study discussing some related oscillation physics and \cite{Rout:2020cxi} for some numerical studies.

In this letter, we will briefly review the standard CP violation picture.
We will then develop the theory for where there is information about CP violation in disappearance measurements.
Finally, we will perform numerical studies indicating the sensitivity to measure $\delta$, and thus determine if CP is violated or not, via disappearance measurements only.

\section{Conventional CP violation picture}
It is true that, consistent with conventional wisdom in the literature, disappearance channels are CP invariant, see e.g.~\cite{Dick:1999ed,T2K:2001wmr,Whisnant:2002fx,Yokomakura:2002av,Nunokawa:2007qh,Akhmedov:2008qt}, under the assumption that CPT is conserved.
That is, by CPT conservation:
\begin{equation}
P(\nu_\alpha\to\nu_\alpha)=P(\bar\nu_\alpha\to\bar\nu_\alpha)\,,
\end{equation}
in vacuum\footnote{The presence of matter modifies neutrino oscillations \cite{Wolfenstein:1977ue} and induces additional apparent CP (or CPT) violation which has been discussed extensively in the literature \cite{Kuo:1987km,Krastev:1988yu,Toshev:1989vz,Toshev:1991ku,Arafune:1996bt,Parke:2000hu,Akhmedov:2001kd,Xing:2013uxa,Petcov:2018zka,Bernabeu:2019npc,Schwetz:2021thj}, but this is easily accounted for and is not the focus of this letter.}.
Thus neutrinos and antineutrinos act the same in vacuum disappearance experiments.

The CP asymmetry, on the other hand, is only nonzero for appearance and is proportional to the Jarlskog invariant $J\equiv s_{12}c_{12}s_{13}c_{13}^2s_{23}c_{23}\sin\delta$ \cite{Jarlskog:1985ht}.
The difference in probabilities for neutrinos and antineutrinos is,
\begin{equation}
P(\nu_\alpha\to\nu_\beta)-P(\bar\nu_\alpha\to\bar\nu_\beta)\simeq\pm8\pi J\frac{\Delta m^2_{21}}{\Delta m^2_{31}}\,,
\label{eq:app CP asymmetry}
\end{equation}
in vacuum near the first oscillation maximum with $\alpha\neq\beta$ where the sign depends on $\alpha$ and $\beta$.
Thus a determination of $J$, which requires measuring all four parameters of the PMNS matrix, indicates how neutrinos and antineutrinos behave differently in appearance oscillation measurements\footnote{If neutrinos have decohered or are in an oscillation averaged regime after traveling many oscillation periods then the term in eq.~\ref{eq:app CP asymmetry} will vanish.}.
Since the Jarlskog invariant shows up in both neutrino mode and antineutrino mode, a measurement of both is not necessary to measure $\sin\delta$, but may help with systematic uncertainties.

New physics such as sterile neutrinos \cite{Acero:2022wqg}, non-standard neutrino interactions \cite{Wolfenstein:1977ue,Proceedings:2019qno,Arguelles:2022tki}, or unitarity violation \cite{Martinez-Soler:2018lcy} could also modify this picture in non-trivial ways by making fully CP conserving scenarios appear CP violating or other such nightmare scenarios.
It may be possible to avoid these scenarios via a combination of experiments at different baselines and energies; see e.g.~\cite{deGouvea:2016pom,Denton:2020uda,Chatterjee:2020kkm}.

There are several other non-conventional means of probing leptonic CP violation.
One is via one-loop corrections to elastic scattering \cite{Sarantakos:1982bp,Marciano:2003eq,Tomalak:2019ibg,Hill:2019xqk} in solar neutrinos which may achieve up to $\sim1\sigma$ sensitivity to CP violation with optimistic experimental assumptions \cite{Brdar:2023ttb}.
This process is still an appearance measurement in that it depends on CP violation via measuring $\nu_e\to\nu_\mu$.
Another is via a one-loop correction to the standard matter effect for neutrinos propagating inside the Sun which leads to an additional matter effect term $5\e{-5}$ times smaller than the standard matter effect \cite{Botella:1986wy} which could lead to a CP violating effect, but it is $\sim10^{-5}$ the size of the standard matter effect and thus beyond the hope even of up-coming experiments \cite{Minakata:1999ze}.

\section{CP violation in disappearance}
\subsection{An understanding via parameter counting}
While it is not directly possible to determine if nature prefers neutrinos or antineutrinos via disappearance measurements alone, it is possible to determine if nature treats neutrinos and antineutrinos the same or differently via measurements of these CP conserving disappearance channels\footnote{It has been pointed out in \cite{Luo:2023xmv} that the CP conserving \emph{parts} of appearance channels could also provide some amount of information about CP violation.}.
That is, disappearance measurements cannot provide information on $\sign(\sin\delta)$ or, equivalently, on $\sign J$, but can constrain $\cos\delta$ and thus potentially rule out CP conserving values of $|\cos\delta|=1$.

The disappearance probability in vacuum for flavor $\alpha$ is
\begin{align}
P(\nu_\alpha\to\nu_\alpha)=1&-4|U_{\alpha1}|^2|U_{\alpha2}|^2\sin^2\Delta_{21}\nonumber\\&-4|U_{\alpha1}|^2|U_{\alpha3}|^2\sin^2\Delta_{31}\nonumber\\&-4|U_{\alpha2}|^2|U_{\alpha3}|^2\sin^2\Delta_{32}\,,
\label{eq:Pdis}
\end{align}
where $\Delta_{ij}=\Delta m^2_{ij}L/4E$ is the kinematic term.

To understand how one can determine if CP is conserved or not, we focus on the four parameters that describe the mixing matrix\footnote{Two additional parameters, the so-called Majorana phases, may also be physical depending on the nature of neutrinos, but their impact in neutrino oscillation experiments is suppressed by $(m_\nu/E_\nu)^2<10^{-14}$ or smaller in oscillations and thus can be safely ignored.}.
We begin by examining the PMNS mixing matrix $U$ in the usual parameterization\footnote{While we are here working in a parameterization dependent framework, the ability to discover CP violation is not artificially enhanced by this and CP violation can be discovered in any unitary parameterization of the mixing matrix because CP violation is a physical effect, see eq.~\ref{eq:J direct} below.} \cite{ParticleDataGroup:2018ovx,Denton:2020igp}:
\begin{widetext}
\begin{equation}
U=
\begin{pmatrix}
c_{13}c_{12}&c_{13}s_{12}&s_{13}e^{-i\delta}\\
-c_{23}s_{12}-s_{23}s_{13}c_{12}e^{i\delta}&c_{23}c_{12}-s_{23}s_{13}s_{12}e^{i\delta}&s_{23}c_{13}\\
s_{23}s_{12}-c_{23}s_{13}c_{12}e^{i\delta}&-s_{23}c_{12}-c_{23}s_{13}s_{12}e^{i\delta}&c_{23}c_{13}
\end{pmatrix}\,.
\label{eq:PMNS}
\end{equation}
\end{widetext}

Since disappearance measurements only constrain absolute values of elements of the PMNS matrix, we notice that the measurements of the first row, the $\nu_e$ row, provide no information about $\delta$.
It would seem that measurements of either the $\nu_\mu$ or $\nu_\tau$ row would provide information about $\delta$, specifically $\cos\delta$, implying that $\nu_e$'s are somehow special and different from the other two flavors\footnote{This apparent difference has been investigated in the literature, see e.g.~\cite{Kimura:2006hy,Akhmedov:2008qt}, but the importance of two simultaneous disappearance measurements has not been realized.}.
In reality, any one row (and any one column) can be made to be ``simple'': only a product of sines, cosines, and $e^{\pm i\delta}$, see, e.g.,~\cite{Denton:2020igp}.
The remaining four elements must always be ``complicated'': the sum or difference of the products of such terms, one of which always contains $e^{\pm i\delta}$.
This provides one means of understanding why two separate disappearance measurements are required to probe CP violation.
That is, if, for example, we had an excellent measurement of $\nu_\mu$ disappearance but not $\nu_e$ or $\nu_\tau$ disappearance, since we could choose to make the $\nu_\mu$ row simple then therefore we cannot learn anything about CP violation.

The absolute value of the complicated elements contains a $\cos\delta$ contribution, which is our means of getting at CP violation.
We also note that in the usual parameterization, the absolute value of the elements in the $\nu_e$ row depends on only two parameters: $\theta_{13}$ and $\theta_{12}$, while the absolute value of the elements in the other two rows each depend on all four parameters in the mixing matrix.

A perfect measurement of a disappearance channel allows for the determination of the coefficients of all three terms, but provides only two constraints on the mixing matrix due to unitarity.
That is, one can always define away one of the $|U_{\alpha i}|^2$ in terms of the other two by $|U_{\alpha1}|^2+|U_{\alpha2}|^2+|U_{\alpha3}|^2=1$.
Thus a perfect measurement of the $\nu_e$ disappearance probability can constrain two parameters which, given how we typically parameterize the mixing matrix, the measurements map onto the parameters $\theta_{13}$ and $\theta_{12}$.
To date, Daya Bay \cite{DayaBay:2018yms} and RENO \cite{RENO:2016ujo} provide excellent constraints on $\theta_{13}$ while KamLAND \cite{KamLAND:2013rgu}, SNO \cite{SNO:2011hxd}, Super-Kamiokande \cite{Super-Kamiokande:2016yck}, and Borexino \cite{BOREXINO:2018ohr} provide good constraints on $\theta_{12}$.
In the future JUNO \cite{JUNO:2022mxj} will measure $\theta_{12}$ with excellent precision.
Thus the $\nu_e$ row is or will be in excellent shape.

For the $\nu_\mu$ row, disappearance measurements provide up to two independent fundamental measurements that map onto four parameters: $\theta_{23}$, $\theta_{13}$, $\theta_{12}$, and $\cos\delta$.
But since $\theta_{13}$ and $\theta_{12}$ are or will be well known, then similar measurements of $\nu_\mu$ disappearance will provide information about $\theta_{23}$ and $\cos\delta$.

Getting directly at the $\Delta_{21}$ oscillations in $\nu_\mu$ disappearance in the same fashion that JUNO does for $\nu_e$ disappearance is extremely challenging given realistic constraints; see the appendix for a discussion of this hypothetical scenario.

We instead focus on leveraging data in planned experiments such as DUNE and HK and a careful spectral measurement to provide information about the beginning of the $\Delta_{21}$ oscillations in a $\Delta_{31}$ and $\Delta_{32}$ dominated regime.
This is similar to the discussed plan for measuring the solar parameters $\Delta m^2_{21}$ and $\theta_{12}$ with Daya Bay data \cite{Forero:2021lax}.
The effect of CP violation thus begins to show up at the low energy side of the $\nu_\mu$ disappearance spectrum, and thus DUNE has an advantage: $\nu_\mu$ experience more oscillations before the $\nu_\mu$ charged-current cross section hits the muon threshold.
While event rates and reconstructions are challenging at lower energies, the effect will impact the rate at which the oscillation maximum decreases where the probability is near one, so there is no probability suppression, which helps the rate.

Since the appearance channel essentially constrains $\sin\delta$ (see eq.~\ref{eq:app CP asymmetry}) while the disappearance channel constrains $\cos\delta$, these two measurements provide key complementary information.
In fact, there will be sign degeneracies in many regions of parameter space of $\delta$ with either only appearance or disappearance.
Moreover, the precision on $\delta$ near $\pi/2$ or $3\pi/2$ is determined by the sensitivity to $\cos\delta$ which comes from this combination of disappearance measurements making this disappearance based measurement crucial for determining the exact value of $\delta$ if we are near $|\sin\delta|=1$ as some data \cite{T2K:2023smv} may be indicating.

\subsection{A direct analytic calculation}
We now present a new direct analytic calculation of CP violation from disappearance measurements.
We find that it is possible to relate the amount of CP violation given by $J$ to the parameters measured in disappearance.
We first note that CP violating effects are proportional to the Jarlskog invariant \cite{Jarlskog:1985ht}
\begin{align}
J&\equiv\Im(U_{\alpha j}^*U_{\beta j}U_{\alpha i}U_{\beta i}^*)
\label{eq:J defn}\\
&=|U_{\alpha j}||U_{\beta j}||U_{\alpha i}||U_{\beta i}|\sin(\phi_{\beta j}+\phi_{\alpha i}-\phi_{\alpha j}-\phi_{\beta i})\,,\nonumber
\end{align}
up to an overall sign, for $\alpha\neq\beta$ and $i\neq j$ where $U_{\alpha i}=|U_{\alpha i}|e^{i\phi_{\alpha i}}$.
Then, starting from a unitarity triangle closure condition along with the row normalization unitarity conditions and some algebra, one finds,
\begin{multline}
J^2=|U_{e2}|^2|U_{\mu2}|^2|U_{e3}|^2|U_{\mu3}|^2\\-\frac14\left(1-|U_{e2}|^2-|U_{\mu2}|^2-|U_{e3}|^2-|U_{\mu3}|^2\right.\\
\left.+|U_{e2}|^2|U_{\mu3}|^2+|U_{e3}|^2|U_{\mu2}|^2\right)^2\,.
\label{eq:J direct}
\end{multline}
This provides an explicit relationship between the parameters measured in disappearance and the amount of CP violation; see the appendix for more details.

We now leverage approximation techniques to theoretically investigate the size of the effect.

\subsection{Analytic approximation}
We now strive to understand exactly how $\cos\delta$, which can provide key information about CP violation, appears in the $\nu_\mu$ disappearance probability in matter in the usual parameterization.
We will see that the matter effect plays a key role in multiple terms of comparable size, making a simple approximation necessarily fairly challenging.

First, we note that the $\Delta_{31}$ and $\Delta_{32}$ terms in eq.~\ref{eq:Pdis} can be approximately combined as mentioned above, see also \cite{Nunokawa:2005nx}.
Thus the $\cos\delta$ dependence in the magnitudes of these two terms will approximately cancel in vacuum, although the matter effect will somewhat change this, see the discussion later in this subsection.
Second, we focus on the $\Delta_{21}$ term.
In vacuum, to first order in $s_{13}$, the term is
\begin{equation}
-4c_{23}^2\left(c_{23}^2s_{12}^2c_{12}^2+s_{23}c_{23}s_{13}\sin2\theta_{12}\cos2\theta_{12}\cos\delta\right)\sin^2\Delta_{21}\,,
\label{eq:21 term}
\end{equation}
where the $\cos\delta$ dependence is numerically $\approx-0.0005\cos\delta$ at $E_{\max}=1.3$ GeV, the energy of the first non-trivial maximum for DUNE.
Thus we would expect that the maxima will be shifted lower for $\cos\delta=1$ and higher for $\cos\delta=-1$.
Third, we include the correction due to the matter effect which significantly changes this.
The matter effect has almost no impact on $\theta_{23}$ or $\delta$ below the atmospheric resonance at $E\simeq11$ GeV \cite{Denton:2016wmg,Xing:2018lob}.
In addition, while $\theta_{13}$, $\Delta m^2_{31}$, and $\Delta m^2_{32}$ do evolve somewhat in matter, they change $\lesssim10\%$ from their vacuum values and the effect can be safely ignored for the $\Delta_{21}$ term.
The solar parameters, $\theta_{12}$ and $\Delta m^2_{21}$, on the other hand, evolve considerably in matter at these energies.
To a sufficient approximation, the matter correction factor for the solar parameters is \cite{Denton:2016wmg,Denton:2018hal}
\begin{equation}
\mathcal S_\odot\simeq\sqrt{(\cos2\theta_{12}-c_{13}^2a/\Delta m^2_{21})^2+\sin^22\theta_{12}}\,,
\end{equation}
where $a=2\sqrt2G_FN_eE$ is the contribution from the matter effect.
At $E_{\max}=1.3$ GeV, $\mathcal S_\odot=3.4$ and provides an excellent approximation to the rescaling of $\Delta m^2_{21}$ in matter.
See the appendix for more discussion of the solar corrections including higher order terms.
We note that past the solar resonance at $E=0.13$~GeV, $\theta_{12}>\pi/4$ and thus $\cos2\theta_{12}<0$ flipping the sign on the $\cos\delta$ dependence.
We can approximate the solar mixing angle in matter by
\begin{equation}
\cos2\theta_{12}\to\frac{\cos2\theta_{12}-c_{13}^2a/\Delta m^2_{21}}{\mathcal S_\odot}\approx-0.96\,,
\end{equation}
which also agrees to excellent precision with the exact answer.
Therefore the second term in the parentheses in eq.~\ref{eq:21 term} changes sign when the matter effect is considered, but the effect is only $0.004\cos\delta$, about half the true effect.
We summarize all the effects comparing the vacuum result to that at HK and at DUNE in table \ref{tab:matter}.
\begin{table}
\centering
\caption{The impact of the matter effect on the $\cos\delta$ dependence of the $\Delta_{21}$ term at the first non-trivial oscillation maximum in vacuum, for HK, and for DUNE.}
\label{tab:matter}
\begin{tabular}{c|c|c|c}
 & Vacuum & HK & DUNE\\\hline
$E$ [GeV] & 0 & 0.3 & 1.3\\
$\mathcal S_\odot$ & 1 & 1.01 & 3.4\\
$s_{13}$ & 0.148 & 0.152 & 0.166\\
$s_{212}c_{212}$ & 0.37 & 0.39 & -0.26\\\hline
Total: & -0.0005 & 0.0005 & 0.004
\end{tabular}
\end{table}

The additional correction comes from the $\Delta m^2_{32}$ term which, in matter, provides an additional $\cos\delta$ dependence of $0.004$ which, when combined with the $\Delta_{21}$ term, adds to $0.008\cos\delta$, in agreement with the exact numerical result.

\begin{figure}
\centering
\includegraphics[width=\columnwidth]{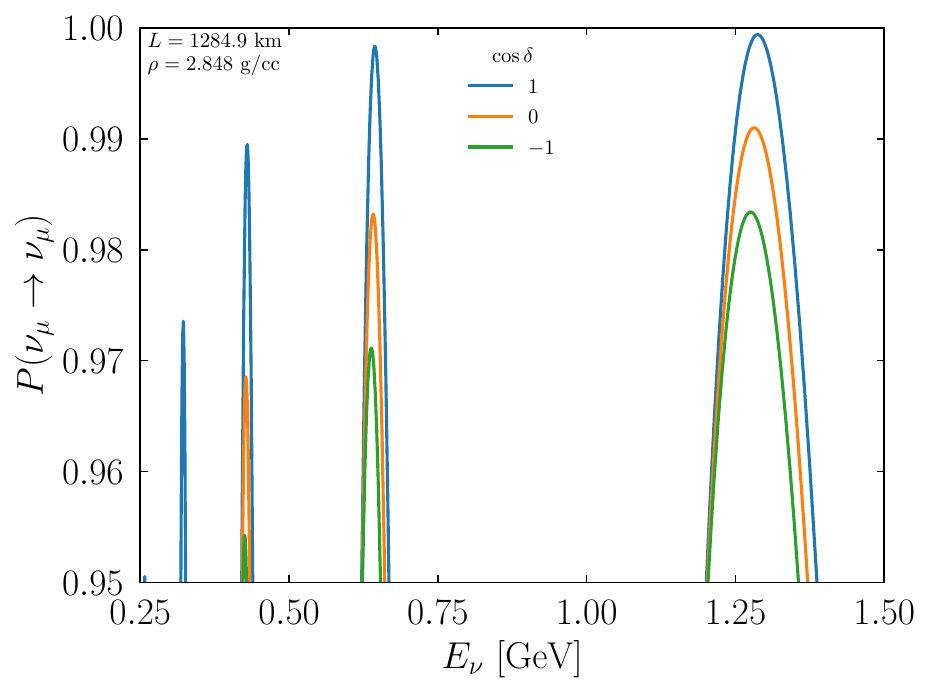}
\caption{The $\nu_\mu$ disappearance probability for DUNE at different values of $\cos\delta$; see the appendix for the same plot for HK.}
\label{fig:probability}
\end{figure}

That is, we expect that the probability in matter should be highest for $\cos\delta=1$ and lowest for $\cos\delta=-1$ varying a total of almost 2\%, as is confirmed numerically in fig.~\ref{fig:probability}.
We have also confirmed that the effect for HK is nearly identical to that in vacuum except for a shift $\cos\delta\to-\cos\delta$; this is because the relevant energy is approximately double the solar resonance in the Earth's crust.
This measurement therefore also provides another indirect test of the matter effect if one is able to compare measurements of $\delta$ between the appearance and the disappearance channels.

For antineutrinos the story is somewhat different.
The value of $\cos\delta$ does not change as $\delta\to-\delta$, so it is the same as for neutrinos.
In matter we note that while $\cos2\wh{\theta_{12}}<0$ for neutrinos for DUNE and HK, it remains positive for antineutrinos, as in vacuum.
Thus the impact on the oscillation maxima for antineutrinos is the same in matter as in vacuum and the probability is comparatively large for $\cos\delta=-1$ and small for $\cos\delta=+1$.
Because of the lower statistics in $\bar\nu_\mu$ mode however, this channel will not contribute as much as the neutrino channel to the total significance for probing $\cos\delta$ and CP violation.

\section{Estimated experimental sensitivities}
\begin{figure}
\centering
\includegraphics[width=\columnwidth]{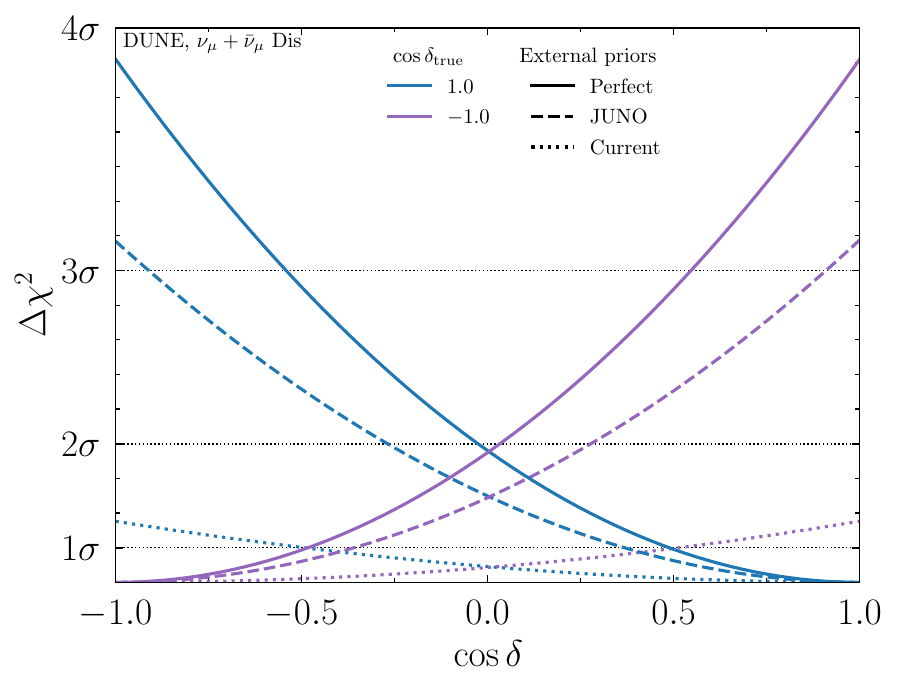}
\caption{The expected sensitivity to $\cos\delta$ using only $\nu_\mu$ and $\bar\nu_\mu$ disappearance from DUNE along with external priors on the other five oscillation parameters from the current precision, the expected improvement with JUNO, or if the other oscillation parameters are known perfectly.}
\label{fig:Delta_cosdelta_individual}
\end{figure}

To numerically quantify the magnitude of the effect given realistic experimental details, we simulate $\nu_\mu$ and $\bar\nu_\mu$ disappearance in DUNE using DUNE's simulation files \cite{DUNE:2021cuw,Huber:2004ka}.
We consider 40 kt fiducial volume and 6.5 years in each neutrino and antineutrino mode with 1.2 MW and 56\% beam uptime.
We consider priors on the five oscillation parameters other than $\delta$ from one of the following:
\begin{enumerate}
\item Our \textbf{current} knowledge of the oscillation parameters \cite{Esteban:2020cvm}.
Since both DUNE and HK will provide better measurements of $\theta_{23}$ and $\Delta m^2_{31}$ than existing data, this is equivalent to using Daya Bay, RENO, KamLAND, and Solar data (all disappearance experiments) to constrain $\theta_{13}$, $\theta_{12}$, and $\Delta m^2_{21}$.
\item The expected improvement on $\theta_{12}$, $\Delta m^2_{21}$, and $\Delta m^2_{31}$ from the inclusion of 6 years of \textbf{JUNO}'s $\bar\nu_e$ disappearance data \cite{JUNO:2022mxj}.
\item The hypothetical scenario with \textbf{perfect} knowledge of all five other oscillation parameters.
\end{enumerate}
We now perform a statistical test to determine DUNE's capability to determine $\cos\delta$ including systematics, efficiency, smearing, and backgrounds as estimated by DUNE \cite{DUNE:2021cuw} and show our results in fig.~\ref{fig:Delta_cosdelta_individual} for each of the three different choices of priors.
For $\cos\delta=\pm1$, the combination of DUNE and JUNO can disfavor $\cos\delta=\mp1$ at $>3\sigma$ and further improvements in JUNO's measurement could reach close to $4\sigma$.
In addition, for $\cos\delta=0$ (CP violating), $|\cos\delta|=1$ (CP conserving) can be disfavored at $1.6\sigma$; see also the appendix.

We have confirmed that there is information about $\cos\delta$ in each of neutrino and antineutrino modes individually, although neutrino mode contributes more to the information due to higher statistics from the larger cross section and lower wrong-sign lepton rates.
In addition, as suggested by the theory discussion above, $\cos\delta$ can also be determined if DUNE was performed in vacuum, although the results would be modified.
Numerous additional numerical results for DUNE as well as HK can be found in the appendix including the impact of runtime.

We also checked the precision with which $\cos\delta$ can be determined.
The $1\sigma$ uncertainty is essentially independent of the true value and is $0.63$ and $0.51$ given external information at the level associated with 6 years of JUNO and perfect knowledge, respectively.

One could also consider $\nu_\mu$ disappearance with atmospheric neutrinos at HK \cite{Hyper-Kamiokande:2018ofw}, IceCube \cite{IceCubeCollaboration:2023wtb}, KM3NeT \cite{KM3Net:2016zxf}, or JUNO \cite{JUNO:2021tll,Suliga:2023pve}, however the expected sensitivity is likely less than that presented here and depends strongly on systematics, see e.g.~\cite{Arguelles:2022hrt} for a discussion including both disappearance and appearance in atmospherics.
Nonetheless, due to the different systematics and timelines it may be useful to consider a fit including atmospheric neutrinos alongside state-of-the-art $\nu_e$ disappearance measurements.

In principle, one could probe $\cos\delta$ with existing disappearance data.
The current status of the data is that the best $\nu_\mu$ disappearance measurements come from NOvA \cite{NOvA:2021nfi} and T2K \cite{T2K:2023smv} and the best $\nu_e$ disappearance measurements come from Daya Bay \cite{DayaBay:2018yms} and KamLAND \cite{KamLAND:2013rgu}.
The $\nu_e$ disappearance data is described by the ``current'' curves in fig.~\ref{fig:Delta_cosdelta_individual} as well as those in the appendix which show that even DUNE or HK can only provide at most $\sim1.4\sigma$ sensitivity to $\cos\delta$; with existing NOvA and T2K data there would not be any significant $\cos\delta$ information at all.

\section{Conclusion}
Determining if CP is violated in the neutrino sector is one of the highest priorities in particle physics.
The best way to do so is with neutrino oscillations in the appearance channels.
As this measurement will face many significant systematic uncertainties, additional means of probing $\delta$ and CP violation will be crucial to ensure robustness.
While disappearance channels are fundamentally CP conserving, we have shown both by counting information in parameters and a direct relationship between $J$ and the $|U_{\alpha i}|^2$'s that disappearance measurements can still provide information about $\delta$, specifically $\cos\delta$, which is sufficient to determine if CP is violated or not.
Nonetheless, it cannot be done with any one disappearance measurement; we require good precision measurements of the disappearance probability of at least two different flavors.

The matter effect affects the details of this story somewhat, but CP violation can be determined in vacuum or matter.
In addition, neutrinos and antineutrinos behave somewhat differently in disappearance due to the matter effect, but neutrino mode alone (or antineutrino mode alone) is sufficient to determine if CP is violated.

In the upcoming generation of experiments, JUNO will measure the $\bar\nu_e$ disappearance probability with unprecedented precision by directly observing all three oscillation frequencies.
Long-baseline experiments like DUNE and HK will measure $\nu_\mu$ disappearance primarily focused on the weighted average of the $\Delta m^2_{31}$ and $\Delta m^2_{32}$ frequencies, but will also detect at a subleading level the $\Delta m^2_{21}$ frequency \cite{Denton:2023zwa}.
This is enough to provide some information about $\delta$.
In particular, we find that DUNE and JUNO combined will be able disfavor some values of $\cos\delta$ at up to $>3\sigma$ depending on the true value.
Since $\nu_\mu$ disappearance has somewhat cleaner and, more importantly, different, systematics from $\nu_e$ appearance in long-baseline measurements at DUNE and HK, this channel will provide a crucial robustness test of CP violation when combined with JUNO data.

\begin{acknowledgments}
The author acknowledges helpful comments from Stephen Parke, Joshua Berger, and the anonymous referee.
The author acknowledges support by the United States Department of Energy under Grant Contract No.~DE-SC0012704.
Data files for the calculations in the letter can be found at \href{https://peterdenton.github.io/Data/CPV\_Dis/index.html}{peterdenton.github.io/Data/CPV\_Dis/index.html}.
\end{acknowledgments}

\makeatletter 
\renewcommand\onecolumngrid{
\do@columngrid{one}{\@ne}%
\def\set@footnotewidth{\onecolumngrid}
\def\footnoterule{\kern-6pt\hrule width 1.5in\kern6pt}%
}

\renewcommand\twocolumngrid{
        \def\footnoterule{
        \dimen@\skip\footins\divide\dimen@\thr@@
        \kern-\dimen@\hrule width.5in\kern\dimen@}
        \do@columngrid{mlt}{\tw@}
}%
\makeatother

\onecolumngrid

\section{Appendix}

\appendix

\section{CP violation discovery sensitivity}
\begin{figure}
\centering
\includegraphics[width=0.49\textwidth]{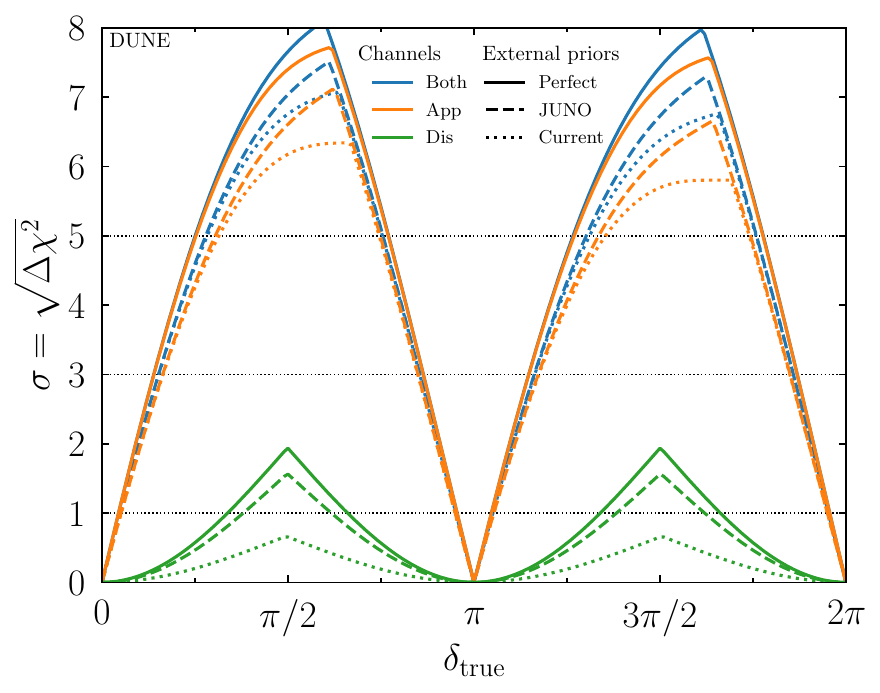}
\includegraphics[width=0.49\textwidth]{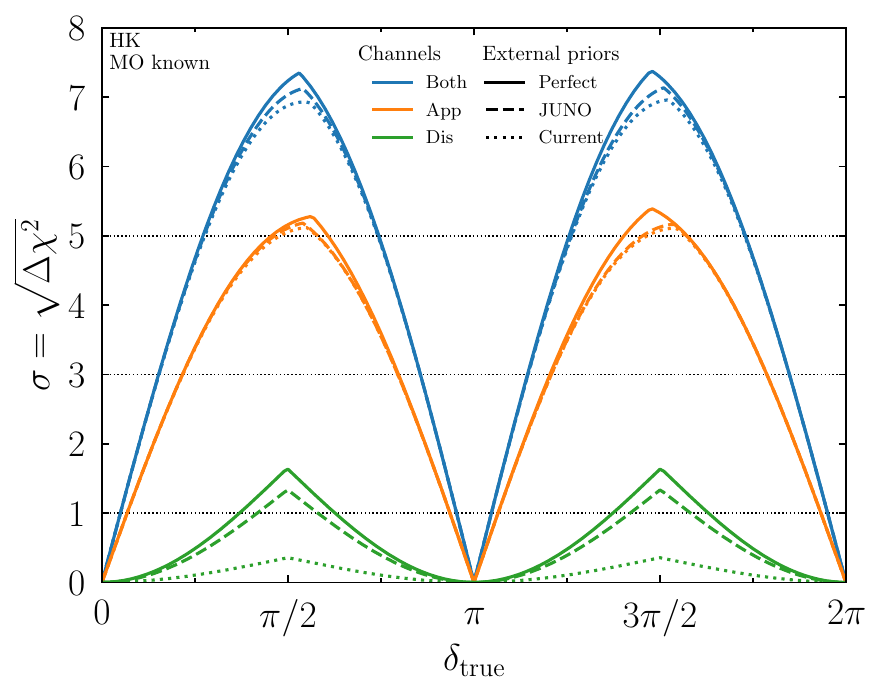}
\caption{The sensitivity of DUNE (left) and HK (right) to disfavor CP conservation, broken down by channel (colors) and the external priors (line styles).}
\label{fig:mcdonalds}
\end{figure}

We now include, for completeness, several additional numerical results including both HK and DUNE.
In fig.~\ref{fig:mcdonalds}, we quantify the expected sensitivity to discover CP violation (that is, ruling out $|\cos\delta|=1$) as a function of the true value of $\delta$ for DUNE (left) and HK (right).
We also consider different choices of external measurements of the other oscillation parameters as in fig.~\ref{fig:Delta_cosdelta_individual}.
The different colors correspond to including both appearance and disappearance (the standard DUNE analysis), only appearance, and only disappearance.
The different line styles correspond to external pulls from the current knowledge of the oscillation parameters, the expected improvements with JUNO, and hypothetical perfect knowledge.
The blue dotted curve (both channels and current knowledge of the oscillation parameters) agrees with DUNE's curve very well.

For HK we assume 1.3 MW, 187 kton fiducial mass, 1:3 neutrino to antineutrino run time ratio, and 10 years of running at 100\% uptime to generally agree with the nominal HK prediction \cite{Hyper-Kamiokande:2018ofw}.
Note that we assume that the mass ordering is known which is relevant for HK and not for DUNE because DUNE will measure it directly at very high significance in the appearance channel.
We find that HK is somewhat less sensitive to discovering CP violation in the disappearance channel than DUNE since the effect is smaller, but the larger statistics mostly compensate for the difference.

We note that the combined combined fit with both appearance and disappearance data yields more information than the naive sum of the $\Delta\chi^2$'s of each separately in the cases with the current or expected JUNO priors due to the fact that $\nu_\mu$ disappearance will provide world leading measurements of $\Delta m^2_{31}$ and $\theta_{23}$, but with perfect knowledge of the other five oscillation parameters, the combined fit is the same as the naive sum of $\Delta\chi^2$'s.

\def\figwidth{0.6\textwidth}

\section{Explicit analytic expression for CP violation from disappearance and unitarity}
We explicitly demonstrate that it is possible to measure and discover CP violation from disappearance measurements alone.
We work in a completely parameterization independent fashion and write: $U_{\alpha i}\equiv|U_{\alpha i}|e^{i\phi_{\alpha i}}$.
We also note that CP violating effects are proportional to the Jarlskog invariant \cite{Jarlskog:1985ht} written in eq.~\ref{eq:J defn}.

We show that the minimal number of measured terms is four and that $|U_{e2}|$, $|U_{e3}|$, $|U_{\mu2}|$, and $|U_{\mu3}|$ suffices, as does any other two-by-two submatrix\footnote{We note that any such measurement actually gives all 9 $|U_{\alpha i}|$ terms by unitarity.}.
First, as shown in eq.~3 in the main text, these can all be measured in disappearance experiment alone, presumably also with $|U_{e1}|$ and $|U_{\mu1}|$, (they are given by unitarity anyway).
Second, in the usual PDG parameterization or a variant thereof \cite{ParticleDataGroup:2018ovx,Denton:2020igp} a measurement of only three parameters provides information about the three real mixing angles.
For example, if experiments measure $|U_{e2}|$ (solar, KamLAND), $|U_{e3}|$ (Daya Bay, RENO), and $|U_{\mu3}|$ (Super-Kamiokande, MINOS, NOvA, T2K, IceCube) then this maps onto knowledge of $\theta_{12}$, $\theta_{13}$, and $\theta_{23}$, respectively, and $\delta$ (and therefore $J$) remains undetermined.
If experiments measure a different set of three parameters, one could perform the same exercise with the same result by parameterizing the matrix differently.
Thus measuring the absolute value of three elements of the PMNS matrix, which has already been done via existing disappearance measurements, is not sufficient to determine if CP is violated.

We can explicitly relate the amplitudes of the oscillation frequencies, which are directly measured, to the $|U_{\alpha i}|$'s discussed in this section.
Specifically, if we write the disappearance probability as
\begin{equation}
P(\nu_\alpha\to\nu_\alpha)=1-4\sum_{i>j}C_{ij}^\alpha\sin^2\Delta_{ij}\,,
\end{equation}
where $C_{ij}^\alpha=|U_{\alpha i}|^2|U_{\alpha j}|^2$ and $C_{ij}^\alpha=C_{ji}^\alpha$.
Then we can directly extract the $|U_{\alpha i}|$'s via
\begin{equation}
|U_{\alpha i}|=\left(\frac{C_{ij}^\alpha C_{ik}^\alpha}{C_{jk}^\alpha}\right)^{1/4}\,,
\end{equation}
for $i$, $j$, $k$ all different.
Note that one of the three $C_{ij}^\alpha$ for a given flavor $\alpha$ can always be written as a function of the other two of the same flavor.

We now assume that $|U_{e2}|$, $|U_{e3}|$, $|U_{\mu2}|$, and $|U_{\mu3}|$ have all been measured.
We begin with the $e-\mu$ triangle closure statement:
\begin{equation}
|U_{e2}||U_{\mu2}|e^{i(\phi_{e2}-\phi_{\mu2})}+|U_{e3}||U_{\mu3}|e^{i(\phi_{e3}-\phi_{\mu3})}=-|U_{e1}||U_{\mu1}|e^{i(\phi_{e1}-\phi_{\mu1})}\,.
\label{eq:triangle}
\end{equation}
We now rewrite $|U_{e1}|$ and $|U_{\mu1}|$ in terms of the parameters we have measured in this scenario via row unitarity,
\begin{align}
|U_{e1}|&=\sqrt{1-|U_{e2}|^2-|U_{e3}|^2}\,,\\
|U_{\mu1}|&=\sqrt{1-|U_{\mu2}|^2-|U_{\mu3}|^2}\,.
\end{align}
We substitute these into eq.~\ref{eq:triangle} above, take the norm-square, and simplify to find
\begin{equation}
2|U_{e2}||U_{\mu2}||U_{e3}||U_{\mu3}|\cos(\phi_{e2}-\phi_{\mu2}-\phi_{e3}+\phi_{\mu3})=
1-|U_{e2}|^2-|U_{\mu2}|^2-|U_{e3}|^2-|U_{\mu3}|^2+|U_{e2}|^2|U_{\mu3}|^2+|U_{e3}|^2|U_{\mu2}|^2\,.
\end{equation}
Given that we will want this to look like the Jarlskog invariant, we rewrite it in terms of sine of the phases by squaring again.
\begin{multline}
4|U_{e2}|^2|U_{\mu2}|^2|U_{e3}|^2|U_{\mu3}|^2\left[1-\sin^2(\phi_{e2}-\phi_{\mu2}-\phi_{e3}+\phi_{\mu3})\right]\\
=\left(1-|U_{e2}|^2-|U_{\mu2}|^2-|U_{e3}|^2-|U_{\mu3}|^2+|U_{e2}|^2|U_{\mu3}|^2+|U_{e3}|^2|U_{\mu2}|^2\right)^2\,.
\end{multline}
Now we take eq.~\ref{eq:J defn} and set $\alpha=e$, $\beta=\mu$, $i=2$, and $j=3$ to get (with the correct sign, which will not matter in this calculation)
\begin{equation}
J=\Im(U_{e3}^*U_{\mu3}U_{e2}U_{\mu2}^*)=|U_{e2}||U_{\mu2}||U_{e3}||U_{\mu3}|\sin(\phi_{e2}-\phi_{\mu2}-\phi_{e3}+\phi_{\mu3})\,.
\end{equation}
Plugging this in and rearranging gives
\begin{equation}
J^2=|U_{e2}|^2|U_{\mu2}|^2|U_{e3}|^2|U_{\mu3}|^2-\frac14\left(1-|U_{e2}|^2-|U_{\mu2}|^2-|U_{e3}|^2-|U_{\mu3}|^2+|U_{e2}|^2|U_{\mu3}|^2+|U_{e3}|^2|U_{\mu2}|^2\right)^2\,.
\end{equation}
This can be naturally adapted for different $2\times2$ submatrices of the PMNS matrix depending on which parameters are most easily measured.

One can also derive this same result using the well known fact that the area of a unitarity triangle is equal to $J/2$ along with Heron's formula:
\begin{align}
\frac J2&=\frac14\sqrt{4|U_{e2}|^2|U_{\mu2}|^2|U_{e3}|^2|U_{\mu3}|^2-\left(|U_{e2}|^2|U_{\mu2}|^2+|U_{e3}|^2|U_{\mu3}|^2-|U_{e1}|^2|U_{\mu1}|^2\right)^2}\,,\\
J^2&=|U_{e2}|^2|U_{\mu2}|^2|U_{e3}|^2|U_{\mu3}|^2-\frac14\left(|U_{e2}|^2|U_{\mu2}|^2+|U_{e3}|^2|U_{\mu3}|^2-|U_{e1}|^2|U_{\mu1}|^2\right)^2\,,\label{eq:J derivation}\\
J^2&=|U_{e2}|^2|U_{\mu2}|^2|U_{e3}|^2|U_{\mu3}|^2-\frac14\left[|U_{e2}|^2|U_{\mu2}|^2+|U_{e3}|^2|U_{\mu3}|^2-\left(1-|U_{e2}|^2-|U_{e3}|^2\right)\left(1-|U_{\mu2}|^2-|U_{\mu3}|^2\right)\right]^2\,,\\
J^2&=|U_{e2}|^2|U_{\mu2}|^2|U_{e3}|^2|U_{\mu3}|^2-\frac14\left(1-|U_{e2}|^2-|U_{e3}|^2-|U_{\mu2}|^2-|U_{\mu3}|^2+|U_{e2}|^2|U_{\mu3}|^2+|U_{e3}|^2|U_{\mu2}|^2\right)^2\,.
\end{align}

It is also easy to check that if one element is zero, say e.g.~$|U_{e3}|=0$, then $J=0$.
From eq.~\ref{eq:J derivation}, this is equivalent to showing that $|U_{e2}|^2|U_{\mu2}|^2=|U_{e1}|^2|U_{\mu}|^2$ which follows from squaring the triangle closure in eq.~\ref{eq:triangle}.

Thus a measurement of only $|U_{e2}|$, $|U_{e3}|$, $|U_{\mu2}|$, and $|U_{\mu3}|$, as is possible from performing good $\nu_e$ and $\nu_\mu$ disappearance experiments, allows for a determination of $J^2$.
Thus it is possible to show that $J\neq0$ -- if nature supports that -- via disappearance measurements.
This approach does not, however, indicate if nature prefers neutrinos or antineutrinos since the sign of $J$ is undetermined and must be measured in a CP violating process such as appearance.

\section{DUNE event rates and regions of interest}
\begin{figure}
\centering
\includegraphics[width=\figwidth]{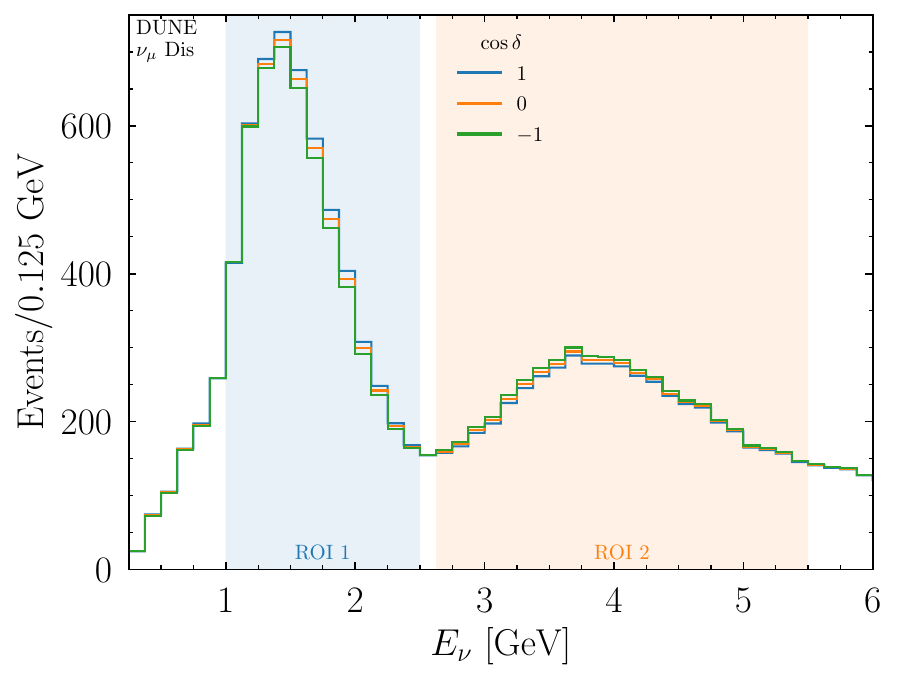}
\caption{The expected $\nu_\mu$ disappearance event rates after 6.5 years in neutrino mode for various values of $\cos\delta$.
We have also defined two regions of interest allowing for a simple statistical understanding of the sensitivity to $\cos\delta$.}
\label{fig:eventrates}
\end{figure}

While the full statistical fits performed for figs.~2 and 3 in the main text contain all of the relevant information, it is useful to gain a physical understanding of where we can expect the effects to appear in the data.
To this end, in fig.~\ref{fig:eventrates} we calculated the expected $\nu_\mu$ event rates after 6.5 years in neutrino mode only including efficiency, smearing, and backgrounds for several key values of $\cos\delta$.
We have defined two interesting regions of interest (ROIs) separated by the local minimum and constrained by half the event rate of each local maximum.
In the lower energy ROI, ROI 1, we see that the event rate increases with $\cos\delta$ while in ROI 2 it decreases with $\cos\delta$.
The statistics in each ROI for the different values of $\delta$ are shown in table \ref{tab:roi} and it is easy to see that each ROI contributes to the statistical test to disfavor CP conservation given $\cos\delta=0$ at the level of $\Delta\chi^2\sim1.4$ based on statistics only leading to a combined estimate of $\sim1.7\sigma$ sensitivity from neutrino mode, close to the correct estimate of $1.6\sigma$ (including neutrino and antineutrino modes) shown in fig.~3 in the main text.
A realistic analysis needs to include uncertainties on the oscillation parameters and systematic uncertainties which will decrease the sensitivity somewhat, but also shape information which will increase it somewhat.

\begin{table}
\centering
\caption{The expected $\nu_\mu$ event rates after 6.5 years of neutrino mode at DUNE as a function of $\cos\delta$ in the two ROIs, see fig.~\ref{fig:eventrates}.}
\label{tab:roi}
\begin{tabular}{c|c|c}
$\cos\delta$&ROI 1&ROI 2\\\hline
1&5506&5038\\
0&5418&5115\\
-1&5334&5193
\end{tabular}
\end{table}

\section{Runtime impact}
\begin{figure}
\centering
\includegraphics[width=\figwidth]{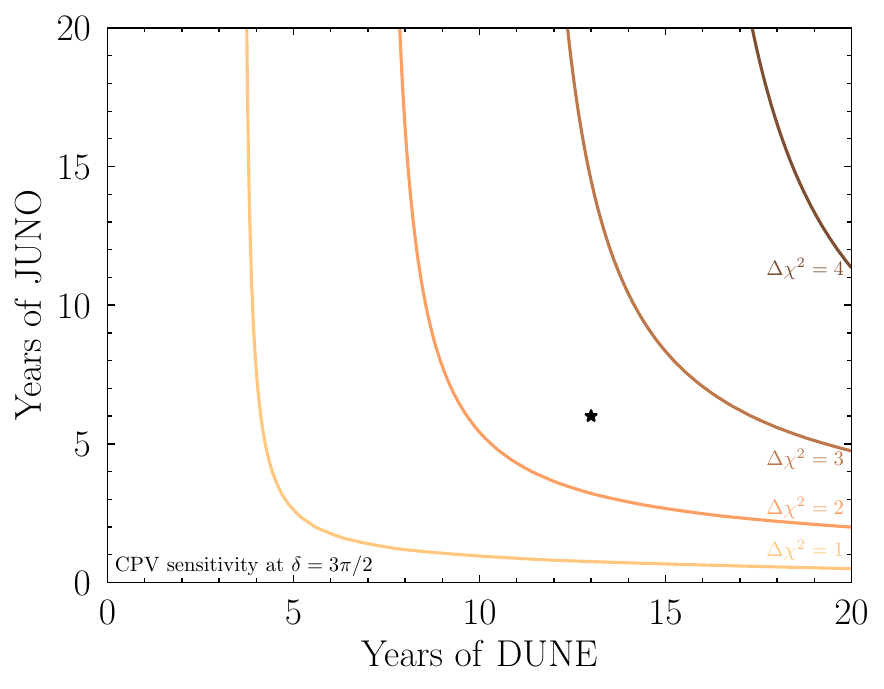}
\caption{The sensitivity to disfavor $\sin\delta=0$ assuming $\delta=3\pi/2$ as a function of DUNE and JUNO's runtime.
The star denotes the benchmark point considered elsewhere in the text.}
\label{fig:runtime contour}
\end{figure}

We also calculate the sensitivity based on the runtime of DUNE and JUNO.
For DUNE we keep the target mass, uptime, and proton power the same and vary the runtime, split evenly between $\nu_\mu$ and $\bar\nu_\mu$.
For JUNO we assume that the precision on the three parameters that they mainly determine, $\Delta m^2_{21}$, $\Delta m^2_{31}$, and $\theta_{12}$, continue to scale as expected with a variance proportional to the sum of run time and a systematic term determined by the numbers in \cite{JUNO:2022mxj}.
We plot the ability to rule out CP conservation assuming $\delta=3\pi/2$; other statistical tests such as ruling out $\cos\delta=1$ assuming $\cos\delta=-1$ (see fig.~2 in the main text) and so on all scale in a similar fashion.
The results are shown in fig.~\ref{fig:runtime contour}.
We see that further improvement beyond the benchmark point requires more time from both experiments.
Instead of additional time, the statistics can also be improved by increasing the beam power (DUNE/HK) or reactor power (JUNO); all experiments have the possibility of seeing such upgrades.
There is a discussion about upgrading the accelerator for DUNE from 1.2 MW to 2.4 MW \cite{DUNE:2022aul,Denton:2022een} and JUNO may benefit from additional nuclear reactors increasing the power from 26.6 GW$_{\rm th}$ to 35.8 GW$_{\rm th}$ \cite{JUNO:2015zny}.
This further highlights the important synergy among long-baseline reactor and accelerator neutrino experiments as improvement in the statistics of either experiment alone will not significantly improve the sensitivity to CP violation, but additional runtime for both will since the effect is not specific to either the $\nu_e$ or the $\nu_\mu$ row.

\section{Hyper-Kamiokande}
\begin{figure}
\centering
\includegraphics[width=\figwidth]{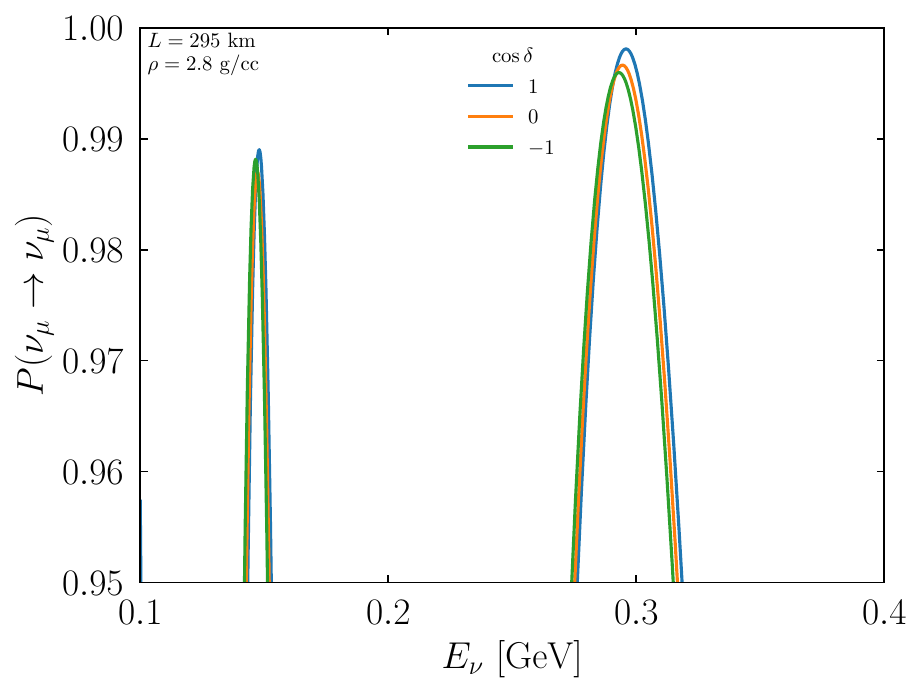}
\caption{The $\nu_\mu$ disappearance probability for HK at different values of $\cos\delta$; see fig.~1 in the main text for the same plot for DUNE.}
\label{fig:probability HK}
\end{figure}

We repeat the same calculations performed in the main text and elsewhere in the appendix, but for HK for completeness.
First we examine the probability itself in fig.~\ref{fig:probability HK} and find that the impact of varying $\cos\delta$ is much smaller than for DUNE, see fig.~1 in the main text, as expected since $\wh{\theta_{12}}$ is closer to $\pi/4$ and thus the leading order $\cos\delta$ dependence in the $\Delta_{21}$ term in eq.~5 in the main text is closer to zero.

\section{Tokai to SURF}
The constraint on $\theta_{13}$, which comes from measuring the quantity $|U_{e3}|^2(|U_{e1}|^2+|U_{e2}|^2)=|U_{e3}|^2(1-|U_{e3}|^2)$, is determined via the kinematic terms $\Delta_{31}$ and $\Delta_{32}$ which has a maximum effect for reactor neutrinos at a baseline of $\sim1.5$ km.
The oscillation\footnote{Solar neutrinos, which largely do not oscillate \cite{Smirnov:2016xzf}, constrain $\simeq|U_{e2}|^2(1-|U_{e3}|^2)$.} constraint on $\theta_{12}$, which comes from measuring the quantity $|U_{e1}|^2|U_{e2}|^2$, is determined via the kinematic term $\Delta_{21}$ which has a maximum effect for reactor neutrinos at a baseline about $\frac{\Delta m^2_{31}}{\Delta m^2_{21}}\simeq30$ times farther of $\sim50$ km.

Disappearance experiments exist that measure $\theta_{23}$ via the quantity $|U_{\mu3}|^2(|U_{\mu1}|^2+|U_{\mu2}|^2)=|U_{\mu3}|^2(1-|U_{\mu3}|^2)$ (combined with information about $\theta_{13}$) at baselines from about 300 km to 10,000 km.
Thus one can get at the second piece of information in the $\nu_\mu$ disappearance picture by performing a $\nu_\mu$ disappearance oscillation experiment at 30 times the baseline for a fixed energy.
This is not feasible for atmospheric neutrinos and one would need to imagine an extremely optimistic configuration from e.g.~Tokai, the accelerator neutrino source for T2K and HK, to SURF, the upcoming far detector location for DUNE, which is a distance of 8235 km, close to the oscillation minimum at 8850 km assuming the same off-axis angle as T2K/HK, although the flux is lower by a factor of almost 800.

We examine the solar $\Delta m^2_{21}$ minimum at a hypothetical very long-baseline $\nu_\mu$ disappearance accelerator neutrino experiment from J-PARC in Tokai, Japan to SURF in Lead, South Dakota, United States at a baseline of 8235 km through the Earth's mantle\footnote{See e.g.~\cite{Kimura:2007mu} for additional discussion of the role of $\cos\delta$ in various $\nu_\mu$ baseline and energy configurations.}.

In vacuum, the oscillation minimum due to $\Delta m^2_{21}$ happens at 0.5 GeV and depends on $\cos\delta$ since the amplitude of the oscillation, to first order in $s_{13}$ is
\begin{equation}
|U_{\mu1}|^2|U_{\mu2}|^2\approx c_{23}^4s_{12}^2c_{12}^2+s_{23}c_{23}^3s_{13}\sin2\theta_{12}\cos2\theta_{12}\cos\delta\,.
\label{eq:Umu1sqUmu2sq}
\end{equation}
While the energies are low, since we are considering oscillations related to $\Delta m^2_{21}$ instead of $\Delta m^2_{31}$, the matter effect will play a role and change the amplitude and shift the location of the minimum.
The matter value of $\theta_{12}$ reaches $\pi/4$ at $2EV_{\rm CC}=\Delta m^2_{21}\cos2\theta_{12}/c_{13}^2$ \cite{Denton:2019yiw,Denton:2018hal} where $V_{\rm CC}=\sqrt2G_FN_e$ is the matter potential and $N_e$ is the electron number density.
For $\rho=4$ g/cc, a typical mantle density, this occurs at $0.09$ GeV

For the frequency, the matter correction is
\begin{equation}
\Delta m^2_{21}\to\wh{\Delta m^2_{21}}\approx\Delta m^2_{21}\mathcal S_\odot\,,
\end{equation}
where the hat denotes a quantity in matter and
\begin{equation}
S_\odot=\sqrt{(\cos2\theta_{12}-c_{13}^2a/\Delta m^2_{21})^2+\sin^22\theta_{12}}\,,
\end{equation}
is the solar correction factor \cite{Denton:2016wmg,Denton:2018hal,Denton:2019yiw}.
We find that the oscillation minimum happens near
\begin{equation}
\wh{\Delta_{21}}=\frac32\pi\,.
\label{eq:Dmsq21a}
\end{equation}
Normally there would be one oscillation minimum at ${\wh{\Delta_{21}}}=\pi/2$, but ${\wh{\Delta_{21}}}>\pi/2$ for all energies at this density and baseline.
This is because at higher energies the matter potential causes $\mathcal S_\odot\propto E$ so $\wh{\Delta m^2_{21}}\propto E$ and thus $\wh{\Delta_{21}}$ is independent of energy.
At lower energies, $\wh{\Delta m^2_{21}}$ reaches a minimum at the solar minimum and continues to increase somewhat to vacuum (roughly proportional to the neutrino energy, but the proximity to the resonance somewhat modifies this dependence), meanwhile the $1/E$ contribution in $\wh{\Delta_{21}}$ ensures that $\wh{\Delta_{21}}$ grows at small energies as well.
Since there is a minimum of $\wh{\Delta_{21}}$ as a function of $E$, then for certain baselines and densities it may be the case that the first (or even higher) oscillation extremum is never reached for any energies.

There is also an additional shift in the location of the minimum due to the prefactor in front of the $\sin^2\Delta_{21}$ term which is proportional to $\sin^22\theta_{12}$ in vacuum which is approximately $\sin^22\theta_{12}/\mathcal S_\odot^2$ in matter, which also depends on the energy.
The shift due to this is small; note that $\theta_{23}$ and $\delta$ don't vary much in matter at all, and the impact of matter on $\Delta m^2_{31}$ and $\theta_{13}$ is small enough at these energies \cite{Denton:2016wmg,Xing:2018lob}.

Thus the location of the minimum is well estimated in this region of parameter space by solving eq.~\ref{eq:Dmsq21a} for the energy,
\begin{equation}
E_{\rm min}\simeq\frac{(\Delta m^2_{21}L)^2/2}{c_{13}^2\cos2\theta_{12}\Delta m^2_{21}L^2V_{\rm CC}+\sqrt{(\Delta m^2_{21}L)^2[(3\pi)^2-(2c_{13}^2\sin2\theta_{12}LV_{\rm CC})^2]}}=0.16{\rm\ GeV}\,.
\label{eq:Emin}
\end{equation}
Since this energy is above $0.09$ GeV, the leading order term proportional to $\cos\delta$ in eq.~\ref{eq:Umu1sqUmu2sq} is nonzero and $\cos\delta=1$ will decrease eq.~\ref{eq:Umu1sqUmu2sq} and thus increase the probability relative to $\cos\delta=0$.
In addition, since the $\cos\delta$ dependence in the next $s_{13}$ order correction depends on $-\cos^2\delta$ and since both orders may be similar due to the $\cos2\theta_{12}$ suppression in the first order term, we see that the probability should be highest for $\cos\delta=1$ and slightly lower, but comparable, for other values of $\cos\delta$.

Next, notice that the size of the large and fast $\Delta_{31}$ and $\Delta_{32}$ amplitudes also depend on $\cos\delta$.
This is because $\Delta_{31}$ and $\Delta_{32}$ are sufficiently different and no longer in phase that they are to be treated separately.
Then we notice that in vacuum through first order in $s_{13}$ the amplitudes are:
\begin{align}
&-4s_{23}^2\left(c_{23}^2s_{12}^2+2s_{23}c_{23}s_{13}s_{12}c_{12}\cos\delta\right)\,,\label{eq:A31}\\
&-4s_{23}^2\left(c_{23}^2c_{12}^2-2s_{23}c_{23}s_{13}s_{12}c_{12}\cos\delta\right)\,,\label{eq:A32}
\end{align}
where the first (second) line is for $\Delta_{31}$ ($\Delta_{32}$).
Since $\wh{s_{12}^2}>\wh{c_{12}^2}$ at these energies, it is the $\Delta_{31}$ term that dominates.
So for $\cos\delta=1$ the dominant amplitude increases in magnitude and the only decrease is in the lesser amplitude, while for $\cos\delta=-1$ the large amplitude is suppressed and the increase in the smaller amplitude brings the two amplitudes closer together.
Then since $\Delta_{31}$ and $\Delta_{32}$ are roughly out of phase at these energies, as expected at the oscillation minimum for $\Delta_{21}$, there is destructive interference when the amplitudes of each phase are close together.

\begin{figure}
\centering
\includegraphics[width=\figwidth]{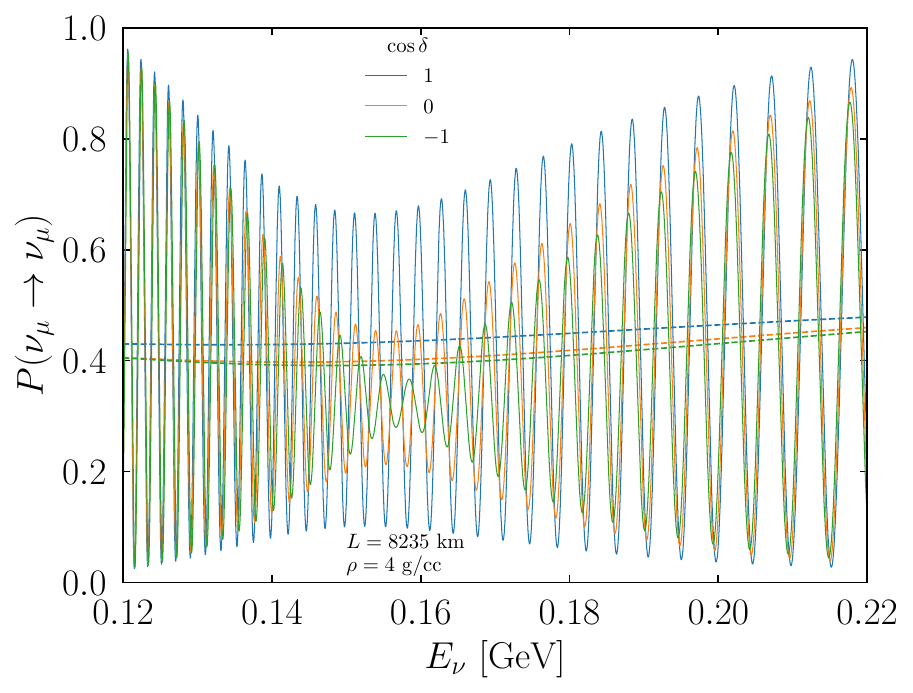}
\caption{The $\nu_\mu$ disappearance probability from Tokai to SURF at different values of $\cos\delta$.
The dashed lines have an additional smearing to show the probability independent of the fast oscillations.}
\label{fig:probability t2s}
\end{figure}

All of these effects discussed above can be seen in fig.~\ref{fig:probability t2s}.
The dashed lines are the same as the solid lines but with a $10\%/\sqrt{E/{\rm GeV}}$ energy resolution smearing, see e.g.~\cite{Friedland:2018vry}.
This shows that even with an optimistic energy resolution, the fast $\Delta_{31}$ and $\Delta_{32}$ oscillations are completely averaged out.
First, the large $\Delta_{21}$ minimum happens at $0.16$ GeV as expected.
Second, we see that the smeared out probability is higher for $\cos\delta=1$ than the other cases which are all similar.
Third, the amplitude of the fast oscillations decreases near the minimum but decrease much more so for $\cos\delta=-1$ than for $\cos\delta=1$.

Experimentally, the low energy of the $\Delta m^2_{21}$ minimum would require the beam to be more off-axis than T2K or HK, which further reduces the flux considerably, and the $\nu_\mu$ CC cross section experiences considerable suppression from the muon mass threshold making an experiment of this nature extremely unfeasible.
An additional issue arises which is one of energy resolution.
Given that the $\Delta m^2_{31}$ and $\Delta m^2_{32}$ oscillations have large amplitude and oscillate $\sim35$ times faster than the larger $\wh{\Delta m^2_{21}}$ oscillation, the effect due to the differing amplitudes of the $\wh{\Delta m^2_{21}}$ oscillations from $\cos\delta$ will be rapidly averaged out unless unrealistically exceptional energy resolution can be achieved.
Thus only the effect described after eq.~\ref{eq:Emin} from eq.~\ref{eq:Umu1sqUmu2sq} could be detectable, not the effect described in eqs.~\ref{eq:A31}-\ref{eq:A32}.
Therefore despite its theoretical interest, this determination of $\cos\delta$ is extremely unfeasible, although theoretically possible.

\vspace*{0.1in}

\twocolumngrid

\bibliography{CPV_Dis}

\end{document}